\documentstyle[eqsecnum,preprint,aps]{revtex}

\def\slashpart{\hbox{$\partial\mkern-10mu/\mkern 1mu$}}
\def\slashA{\hbox{$A\mkern-10mu/\mkern 1mu$}}
\def\slashB{\hbox{$B\mkern-12mu/\mkern 2mu$}}
\def\slashp{\hbox{$p\mkern-8mu/\mkern 0mu$}}
\def\slashk{\hbox{$k\mkern-9mu/\mkern 0mu$}}
\def\slashq{\hbox{$q\mkern-9mu/\mkern 0mu$}}

\begin{document}
\input FEYNMAN
\draft

\title{Abelian Anomalies in Nonlocal Regularization}
\author{M. A. Clayton\thanks{E-mail address: clayton@medb.physics.utoronto.ca},
L. Demopoulos and J. W. Moffat}
\address{University of Toronto\\
Toronto, On., Canada\\
M5S 1A7}
\date{\today}
\maketitle

\begin{abstract}

Nonlocal regularization of QED is shown to possess an axial anomaly
of
the same form as other regularization schemes.
The Noether current is explicitly constructed and the symmetries
are shown to be violated, whereas the identities
constructed when one properly considers the contribution from the
path integral measure are respected.
We also discuss the barrier to quantizing the fully gauged chiral
invariant theory, and consequences.
\end{abstract}

\hbox{UTPT-93-05}
\narrowtext

\section*{Introduction}

A scheme for the nonlocal regularization of gauge theories has
recently been introduced \cite{Moff-Wood,KW-YM,KW-2L} which, aside
from preserving the physical aspects of gauge invariance, is also
finite, Poincar\'e invariant, and perturbatively unitary, without
changing the dimension of spacetime or altering the pole structure
(particle content).
It also avoids the ambiguities associated with defining
$\gamma^5$ in fractional spacetimes, and therefore provides an
ideal stage for examining chiral anomalies.
An important aspect of nonlocal regularization is that the Lagrangian
is presented in regulated form at the beginning.
In local field theory, the regularization method is invoked at the level of
the calculation of diagrams, which leads to many of the inherent problems
and ambiguities.
In this paper, we shall examine the ABJ anomaly \cite{Adler} in the
nonlocal regularization of QED to determine how the anomaly will
manifest
itself.

Because the consequences of the chiral anomaly are well known (fermion
doubling to retain renormalizability, the correct  rate for
$\pi^0\rightarrow\gamma\gamma$), it would be truly surprising for
the nonlocal regularization to bypass it consistently, while
retaining any sort of reasonable local limit.
Here we will show that nonlocal QED produces an anomaly in the
perturbative expansion, and that when the Ward identities are
correctly derived by considering the Jacobian of the measure under
a local axial transformation, they are satisfied and a consistent
local limit is obtained.

In section 2, we develop conventions and briefly review results
from local QED. Section 3 develops an equivalent classical theory
as a precursor to nonlocal quantization, following which the
quantization is performed and Ward identities and relevant loop
corrections are obtained in section 4. In section 5, we discuss
the failed attempt to gauge the full chiral invariance.

\section{Local QED}
\label{local}

We begin by briefly reviewing local QED in order to establish
conventions and the method we will follow in developing the
anomaly in the nonlocal theory.
The standard Lagrangian for local QED is written as
\begin{eqnarray}
L&=&{\bar\psi}(i\slashpart -m)\psi -\frac{1}{4e^2}F^2-
{\bar\psi}\slashA\psi\nonumber \\
&\equiv&{\bar\psi} S^{-1}\psi+\frac{1}{2e^2}A^\mu
D^{-1}_{\mu\nu}A^\nu -{\bar\psi}\slashA\psi,
\end{eqnarray}
which possesses the infinitesimal gauge invariance:
\begin{equation}
\delta A^\mu=\theta^{,\mu},\quad\delta\psi=-
i\theta\psi,
\end{equation}
giving rise to the conserved vector current
$\bar{\psi}\gamma^\mu\psi$.
We have introduced the inverse propagators into the Lagrangian in
order to clarify notation later and in doing so we are assuming
that the trivial gauge fixing procedure has been performed on the
photon.
In the chiral limit, this local Lagrangian also has a global axial invariance
$\delta\psi=-i\omega\gamma^5\psi$ ($\omega =const.$), and the
associated Noether current:
\begin{equation}\label{locax}
J^\mu_5= \bar{\psi}\gamma^\mu\gamma^5 \psi,
\end{equation}
is classically conserved.
When fermion masses are present the equations of motion give:
\begin{equation}\label{div5}
\partial_\mu J^\mu_5=2imJ_5.
\end{equation}

It is well known that this current is no longer conserved when
one quantizes the theory, and QED is then said to have an anomaly.
(This persists when m$\neq$0 as the current does not obey the
classical equations of motion (\ref{div5}).)
This result is easily seen by computing the second order correction
to the axial current coupling to two photons \cite{itch}.
It was realized by Fujikawa \cite{Fujikawa}, that one
could understand the anomaly in the path integral by considering
carefully the transformation properties of the properly regulated
measure, and although the anomaly cannot be removed from the
theory, it is possible to generate consistent Ward identities by
considering the Jacobian of the following local, infinitesimal
change of fermion variables\cite{Ramond}:
\begin{equation}\label{loc1}
\delta\psi(x)=-i\omega(x)\gamma^5\psi(x).
\end{equation}
Writing the generating functional:
\begin{eqnarray}\label{genfun}
Z[S_\mu,\bar{\eta},\eta]&=&\int d\mu[A^\mu,\psi,\bar{\psi}]\nonumber \\
&\times&exp[i\int d^4x(L+S_\mu A^\mu+\bar{\eta}\psi+\bar{\psi}\eta)],
\end{eqnarray}
one transforms the fermionic degrees of freedom as in
Eq. (\ref{loc1}), and generates Ward identities for the axial
current by setting the infinitesimal variation of the generating
functional to zero:
\begin{equation}\label{cond}
\frac{\delta}{\delta\omega}Z[S_\mu,\bar{\eta},\eta]\vline_{\omega
=0}=0.
\end{equation}
To complete this process, one has to carefully consider the
definition of the path integral with regards to the measure and
some form of regulation procedure \cite{Fujikawa}.
One then finds that the naively trivial Jacobian actually produces
\begin{eqnarray}
d\mu[\psi,\bar{\psi}]\rightarrow d\mu[\psi,\bar{\psi}]
exp[-\frac{i}{8\pi^2}\int dx\omega
\tilde{F}^{\mu\nu}F_{\mu\nu}dx],
\end{eqnarray}
where we have introduced the dual field strength as
$\tilde{F}^{\mu\nu}=\frac{1}{2}\epsilon^{\mu\nu\alpha\beta}
F_{\alpha\beta}$.
Combined with the transformation properties of the Lagrangian:
\begin{equation}\label{deltaL}
\delta L=2im\omega J_5+J^\mu_5\partial_\mu\omega,
\end{equation}
this non-invariance of the measure leads to the anomalous
weak operator conservation law
\begin{equation}\label{axial}
J^\mu_{5,\mu}=2imJ_5-\frac{1}{8\pi^2}\tilde{F}^{\mu\nu}
F_{\mu\nu}.
\end{equation}

One can then perform the simplest quantum correction to the axial
current in the presence of two external photons via fig. \ref{triangle}, and
after imposing vector current conservation and taking into account the form
of the pseudoscalar ($\gamma^5$) and pseudovector ($-i\gamma^\mu\gamma^5$)
couplings, one finds
\begin{equation}\label{llim}
p_\mu\Gamma^{\mu\alpha\beta}_5=2m\Gamma^{\alpha\beta}_5-
\frac{i}{2\pi^2}\epsilon^{\mu\nu\alpha\beta}
q_{1\mu}q_{2\nu},
\end{equation}
consistent with (\ref{axial}).

At this point, it is worth stressing that in local theories it
does not make any difference whether one includes an explicit axial
coupling into the Lagrangian, since the structure of the theory
does not change.
In the case of nonlocal regularization, we will find that the axial
coupling must be present in the beginning in order to generate a
classical theory that respects (vector) current conservation.

\section{Nonlocal QED}
\subsection{Shadow Field Formalism}

We will begin by introducing two types of propagators, smeared by
an entire function that possesses strong convergence properties
in the Euclidean regime:
\begin{eqnarray}
\hat{S}(p)&=&E^2(p)S(p)\nonumber \\
&=&-\int_1^\infty\frac{dx}{\Lambda^2}
exp(x\frac{p^2-m^2}{\Lambda^2})(\slashp +m)\nonumber \\
\bar{S}(p)&=&(1-E^2(p))S(p)\nonumber \\
&=&-\int_0^1\frac{dx}{\Lambda^2}
exp(x\frac{p^2-m^2}{\Lambda^2})(\slashp+m),
\end{eqnarray}
where
\begin{equation}
E(p^2)=exp(\frac{p^2-m^2}{2\Lambda^2}).
\end{equation}

Here, we have given their Schwinger parameterized form.
This is especially useful when calculating diagrams, since one
merely writes the local graph in Schwinger parameter form, and
restricts the range of parameter integrals appropriate for the
process in question \cite{KW-2L}.
For example, when one calculates single loop graphs, the unit
hypercube is removed from the volume of integration.
We now construct the auxiliary Lagrangian:
\begin{eqnarray}\label{aux1}
L_{Sh}&=&{\bar\psi}\hat{S}^{-1}\psi+{\bar\phi}{\bar S}^{-
1}\phi\nonumber \\
&+&\frac{1}{2e^2}A^\mu D^{-1}_{\mu\nu}A^\nu-
({\bar\psi}+{\bar\phi})\slashA(\psi +\phi).
\end{eqnarray}
This particular choice of Lagrangian corresponds to a nonlocal
regularization of QED, in which the classical theory retains the
smearing on the internal photon lines, and internal fermion lines
are `localized'.
It is a simple matter to convince oneself of this, since we have by
construction a Lagrangian that generates tree diagrams in which
every $\psi$ coupling has a $\phi$ coupling corresponding to it,
and so every tree process will consist of two separate graphs for
each fermion line, one with a `hatted' propagator and one with a `barred'
one, the two adding to give the local propagator.
This guarantees decoupling of longitudinal photons at the classical
level \cite{Moff-Wood}.
(We could have `localized' the entire theory by introducing
`shadow' fields for the photon as well as the fermions, in which
case the classical theory would be identical to the local theory,
and we would have a viable physical theory of QED with a
fundamental scale.)
\par
It is simple to see that (\ref{aux1}) is invariant under:
\begin{eqnarray}
\delta A^\mu&=&\theta^{,\mu}\nonumber \\
\delta\psi&=&-iE^2\theta(\psi+\phi),\nonumber \\
\delta\phi&=&-i(1-E^2)\theta(\psi+\phi).\nonumber
\end{eqnarray}
The conserved Noether current (generalizing the local vector
current) is given by
\begin{equation}
J^\mu=(\bar{\psi}+\bar{\phi})\gamma^\mu(\psi+\phi).
\end{equation}
The shadow fields are introduced merely as a device to
generate the nonlocal action and symmetries in a compact form.
They do not have a pole in their propagator and hence are not
propagating degrees of freedom, and they should not be included in
asymptotic states.
To generate the action in terms of physical fields alone, we must
integrate them out of the action by forcing them to obey their
equations of motion.
We have
\begin{equation}
\phi=\bar{S}\slashA(\psi +\phi)=(1-\bar{S}\slashA)^{-1}
\bar{S}\slashA\psi,
\end{equation}
and the Lagrangian, gauge transformations and Noether current are
given by:
\begin{eqnarray}
L&=&\bar{\psi}\hat{S}^{-1}\psi+\frac{1}{2e^2}A^\mu\hat{D}^{-1}_{\mu\nu}
A^\nu\nonumber \\
&-&\bar{\psi}(1-\slashA\bar{S})^{-1}\slashA(1-
\bar{S}\slashA)^{-1}\psi \nonumber \\
&+&\bar{\psi}\slashA\bar{S}(1-\slashA\bar{S})^{-1}\bar{S}^{-1}(1-
\bar{S}\slashA)^{-1}\bar{S}\slashA\psi\nonumber \\
&=&\bar{\psi}\hat{S}^{-1}\psi+\frac{1}{2e^2}A^\mu\hat{D}^{-
1}_{\mu\nu}A^\nu-\bar{\psi}\slashA(1-\bar{S}\slashA)^{-1}\psi,
\end{eqnarray}
\begin{equation}
\delta\psi=-iE^2\theta(1-\bar{S}\slashA)^{-1}\psi,
\end{equation}
\begin{equation}
J^\mu=\bar{\psi}(1-\slashA\bar{S})^{-1}\gamma^\mu(1-
\bar{S}\slashA)^{-1}\psi.
\end{equation}
These results reproduce the classical theory described in
\cite{Moff-Wood} (up to a rescaling of the electromagnetic field strength
$A\rightarrow eA$, and a sign convention on the coupling).

We shall now discuss the effects of an axial coupling on
QED, and explore the consequences of loop effects on the classical
equations of motion.
But we must first consider the effects of
an axial coupling in this regularization scheme.
Consider the nonlocal regularized version of the global axial
invariance:
\begin{eqnarray}\label{nlocax}
\delta\psi&=&-iE^2\omega\gamma^5(\psi+\phi),\nonumber \\
\delta\phi&=&-i(1-E^2)\omega\gamma^5(\psi+\phi).
\end{eqnarray}
If we consider graphs containing corresponding current insertions-even at the
classical level-we no longer have current
conservation (axial nor vector), and the longitudinal degree of
freedom of the photon therefore does not decouple.
This can be seen by computing any tree process with two or more
axial current insertions, and coupling in an external longitudinal
photon (or axial boson), and observing that these tree processes
are not `localized'.
No axial couplings occur in the shadow field equations of motion
and, therefore, any diagram with two adjacent axial insertions will not
receive the `barred' fermion propagator contribution in the full
nonlocal Lagrangian.
This is trivially due to the fact that we did not include the axial
coupling in the construction of the nonlocal Lagrangian, and
consequently have not guaranteed decoupling and gauge
invariance in its presence.

\subsection{Nonlocal QED with Axial Couplings}

We now generate a nonlocal Lagrangian that respects
current conservation and decoupling at tree level in the presence
of axial couplings, and therefore contains the physics of QED.
In order to do this we begin again at the local level with
\begin{eqnarray}
L&=&{\bar\psi} S^{-1}\psi+\frac{1}{2e^2}A^\mu
D^{-1}_{\mu\nu}A^\nu\nonumber \\
&-&{\bar\psi}\slashA\psi
-{\bar\psi}\slashB\gamma^5\psi-iC\bar{\psi}\gamma^5\psi,
\end{eqnarray}
where we have included an axial vector coupling to some field(s)
$B^\mu$ and (for convenience) a pseudoscalar coupling to a
pseudoscalar field C (neither field having U(1) vector
transformation properties nor additional photon interactions).
We then repeat the shadow field construction to generate the
Lagrangian:
\begin{eqnarray}
L_{Sh}&=&{\bar\psi}\hat{S}^{-1}\psi+{\bar\phi}{\bar S}^{-
1}\phi+\frac{1}{2e^2}A^\mu D^{-1}_{\mu\nu}A^\nu \nonumber \\
&-&({\bar\psi}+{\bar\phi})\slashA(\psi +\phi)
-({\bar\psi}+{\bar\phi})\slashB\gamma^5(\psi +\phi)\nonumber \\
&-&iC(\bar{\psi}+\bar{\phi})\gamma^5(\psi+\phi),
\end{eqnarray}
and integrate out the shadow fields (defining
$\Gamma=\slashA+\slashB\gamma^5+iC\gamma^5$):
\begin{equation}
\phi=(1-\bar{S}\Gamma)^{-1}\bar{S}\Gamma\psi,\nonumber \\
\end{equation}
\begin{equation}\label{nlocact}
L_{NL}=\bar{\psi}\hat{S}^{-1}\psi+\frac{1}{2e^2}A^\mu\hat{D}^{-1}
_{\mu\nu}A^\nu-\bar{\psi}\Gamma(1-\bar{S}\Gamma)^{-1}\psi.
\end{equation}
Then, invariance under the transformation:
\begin{equation}\label{delpsi}
\delta\psi=-ieE^2\theta(1-\bar{S}\Gamma)^{-1}\psi
\end{equation}
is guaranteed.
The shadow field equations give the following classically conserved
vector and axial vector Noether currents:
\begin{eqnarray}\label{nlocurrents}
J^\mu&=&(\bar{\psi}+\bar{\phi})\gamma^\mu(\psi+\phi)\nonumber \\
&=&\bar{\psi}(1-\Gamma\bar{S})^{-1}\gamma^\mu(1-\bar{S}\Gamma)^{-1}\psi,
\nonumber \\
J^\mu_5&=&(\bar{\psi}+\bar{\phi})\gamma^5\gamma^\mu(\psi+\phi)\nonumber \\
&=&\bar{\psi}(1-\Gamma\bar{S})^{-1}\gamma^5\gamma^\mu(1-\bar{S}
\Gamma)^{-1}\psi.
\end{eqnarray}
Let us also define the pseudoscalar density:
\begin{eqnarray}
J_5&=&(\bar{\psi}+\bar{\phi})\gamma^5(\psi+\phi)\nonumber \\
&=&\bar{\psi}(1-\Gamma\bar{S})^{-1}\gamma^5(1-\bar{S} \Gamma)^{-1}\psi.
\end{eqnarray}
The variation of the generated Lagrangian under
(\ref{nlocax}) gives the same result as in the local case:
\begin{equation}
\delta L=2i\omega m({\bar\psi}+{\bar\phi})\gamma^5(\psi
+\phi)\equiv 2i\omega mJ_5,
\end{equation}
and the equations of motion also give (\ref{div5}).
We have thus explicitly seen one major difference between local
theories and their nonlocal `extensions', namely, that it is
necessary to consider the effect of all of the interaction terms
when constructing the nonlocal Lagrangian, otherwise we cannot
guarantee that the classical action will display the symmetries of
the unregulated theory.
\section{Quantizing the theory}

\subsection{Generating the Measure}
\label{measure}

We now wish to quantize the theory described by (\ref{nlocact}) in
the path integral formalism, and see to what extent classical
current conservation is respected.
Doing this requires finding an invariant measure that respects the
full nonlocal gauge invariance described by (\ref{delpsi}), since
the trivial measure is no longer invariant.
We therefore need a method to generate an invariant measure in
order to consistently quantize the theory and retain the nonlocal
invariance in the quantum regime, thereby guaranteeing
decoupling.
The simplest way to do this is to derive conditions on the
invariant measure by considering how the trivial measure transforms
under the nonlocal regularization gauge transformations, and
requiring that an additional measure contribution compensates.

We write the full invariant measure as the product of the
trivial measure and an exponentiated action term:
\begin{equation}
\mu_{inv}[\phi]=D[\phi]exp(iS_{meas}[\phi]),
\end{equation}
and then perform a gauge transformation and require that the full
measure be invariant. Functionally integrating to derive the
measure yields:
\begin{equation}
\delta\mu_{inv}[\phi]=\mu_{inv}[\phi](i\delta
S_{meas}+Tr[\frac{\partial}{\partial\phi}\delta\phi])=0.
\end{equation}
The trace appears as the only surviving diagonal terms of the
Jacobian determinant of the infinitesimal transformation (when
dealing with fermions, the grassman derivatives will produce the
necessary extra minus sign that corresponds to the inverse
determinant).
We then have
\begin{equation}
\delta S_{meas}=iTr[\frac{\partial}{\partial\phi}\delta\phi].
\end{equation}

This procedure only determines the measure up to gauge invariant
terms, but we feel that any such terms have no place in the
measure, since they properly belong in the Lagrangian and we
generate only the minimal measure necessary for invariance.
We also note that there is, in general, a fair degree of
arbitrariness in constructing the form of the measure.
Each choice produces an equivalent theory, and corresponds to
maintaining a different gauge condition after radiative
corrections.
The measure is also constrained to be an entire function of the
4-momentum invariants for the particular process, ensuring that no
additional degrees of freedom become excited in the quantum regime.
We shall see later on that it is possible to quantize the full
chiral invariance,  if we are prepared to give up this constraint
or add additional particle content at the classical level.

Specifically, the nontrivial contributions to the measure at second
order come from:
\begin{equation}
\delta\psi=-iE^2\theta\bar{S}\slashA\psi,
\end{equation}
and, as given in the original paper\cite{Moff-Wood}, produce the
necessary contribution to vacuum polarization in order to satisfy
the Ward identity and keep the photon transverse:
\begin{equation}\label{vac}
S_m=-\frac{\Lambda^2}{4\pi^2}\int\frac{dpdq}{(2\pi)^4}(2\pi)^4\delta^4(p+q)
M_v(q)A^\mu(p)A_\mu(q),
\end{equation}
\begin{equation}
M_v(q)=\int_0^\frac{1}{2}dt(1-t)exp(t\frac{q^2}{\Lambda^2}-
\frac{1}{1-t}\frac{m^2}{\Lambda^2}),
\end{equation}
(where the massless limit will be denoted by $M_v^0$).

The new piece comes at third order ($BA^2$) and, as we shall see,
is required for decoupling of the longitudinal photon from the
induced A-V-V interaction.
It comes from considering the transformation of the trivial
measure under
\begin{equation}
\delta\psi=-iE^2\theta\bar{S}\Gamma\bar{S}\Gamma\psi,
\end{equation}
giving
\widetext
\begin{eqnarray}\label{Smeas}
S_{meas}&=&\frac{-i}{2\pi^2}\int\frac{dpdq_1dq_2}{(2\pi)^6}(2\pi)^4
\delta^4(p+q_1+q_2)\nonumber \\
&&\epsilon^{\mu\nu\alpha\beta}B_\mu(p)q_{1\nu}A_\alpha(q_1)A_\beta(q_2)
[M_a(p;q_1,q_2)+M_a(q_1;p,q_2)]
\end{eqnarray}
where
\begin{eqnarray}\label{Mdelt}
M_a(p;q_1,q_2)&=&\int_0^1\int_0^1\frac{dxdy}{(1+x+y)^3}\nonumber \\
&\times&exp[\frac{xy}{1+x+y}\frac{p^2}{\Lambda^2}
+\frac{x}{1+x+y}\frac{q_1^2}{\Lambda^2}
+\frac{y}{1+x+y}\frac{q_2^2}{\Lambda^2}-
(1+x+y)\frac{m^2}{\Lambda^2}]
\end{eqnarray}
(where again the case of massless fermions will be denoted by $M_a^0$).
This term in the action produces a Feynman rule:
\begin{equation}\label{Gams}
-i\Gamma^{\mu\alpha\beta}_{5 meas}=-\frac{1}{2\pi^2}
\epsilon^{\mu\nu\alpha\beta}
[q_{2\nu}(M_a(p;q_1,q_2)+M_a(q_2;p,q_1))
-q_{1\nu}(M_a(p;q_1,q_2)+M_a(q_1;p,q_2))].
\end{equation}
\narrowtext

We now have the measure necessary to preserve the vector
invariance in the QED sector (to third order) in the presence of
axial interactions.
As we will see in the next section, this measure will not
help preserve axial vector current conservation and the theory has
an anomaly.

\subsection{Radiative Corrections to the Axial Current}

When we calculate the triangle graph, we will denote the sum over
graphs (with appropriate factors of $E^2$ and $(1-E^2)$) by
$\sum_{R^n_1}$ ($n=3$), where we exclude the graph that
corresponds to three `barred' fermion lines that would cause the
loop to be fully localized, hence divergent.
Normally regularization of loop corrections is performed at the diagram
level, and in the case of the anomaly, the linear divergence leads
to momentum routing ambiguities, or problems defining $\gamma^5$ in
dimensional regularization.
Here, the action is regulated, and it produces loop integrals that
are strongly convergent, without leaving four dimensions.

\widetext
Writing the amplitudes in Fig. \ref{triangle} as
\begin{eqnarray}
-iA^{\mu\alpha\beta}&=&-Tr\int\frac{d^4k}{(2\pi)^4}\sum_{R1^3}
\gamma^\mu\gamma^5S(k)\gamma^\alpha
S(k-q_1)\gamma^\beta S(p+k) \\
-iB^{\mu\alpha\beta}&=&-Tr\int\frac{d^4k}{(2\pi)^4}\sum_{R1^3}
\gamma^\mu\gamma^5S(-k-p)\gamma^\beta
S(q_1-k)\gamma^\alpha S(-k),
\end{eqnarray}
we first check that the longitudinal photon decouples by dotting
$q_{1\alpha}$ into this and writing
$\slashq_1=-(\slashk-\slashq_1-m)+(\slashk-m)$ and
simplifying:
\begin{eqnarray}
-iq_{1\alpha}\Gamma_5^{\mu\alpha\beta}&=&-
iq_{1\alpha}(A^{\mu\alpha\beta}+B^{\mu\alpha\beta})\nonumber \\
&=&-8i\epsilon^{\mu\nu\alpha\beta}\int\frac{d^4k}{(2\pi)^4}
\sum_{R1^3}[\frac{k_\alpha p_\nu}{(k^2-m^2)((p+k)^2-m^2)}-
\frac{k_\alpha p_\nu-q_{1\alpha}p_\nu-q_{1\alpha}k_\nu}{((k-q_1)^2-m^2)
((p+k)^2-m^2)}] \nonumber \\
&=&\frac{1}{2\pi^2}\epsilon^{\mu\nu\alpha\beta}q_{1\alpha}
q_{2\nu}[M_a(p;q_1,q_2)+M_a(q_2;p,q_1)],
\end{eqnarray}
(where we have used
$Tr[\gamma^5\gamma^\alpha\gamma^\beta\gamma^\mu\gamma^\nu]=-
4i\epsilon^{\alpha\beta\mu\nu}$ with $\epsilon^{0123}=+1$) which is
trivially seen to be cancelled by the contribution from the measure
term (\ref{Gams}).

In terms of the axial current conservation, dotting $p_\mu$ into
the same diagram gives (after writing
$\slashp=(\slashp+\slashk-m)-(\slashk+m)+2m$ which allows us to
reduce the traces and immediately recognize the axial coupling term
that satisfies the classical Ward identity, separating the
anomalous term):
\begin{eqnarray}\label{npcalc}
-ip_\mu\Gamma_5^{\mu\alpha\beta}&=&-ip_\mu(A^{\mu\alpha\beta}+
B^{\mu\alpha\beta})\nonumber \\
&=&-2im\Gamma_5^{\alpha\beta}\nonumber \\
&-&8i\epsilon^{\mu\nu\alpha\beta}\int\frac{d^4k}{(2\pi)^4}\sum_{R1^3}
[\frac{k_\mu q_{1\nu}}{(k^2-m^2)((p+k)^2-m^2)}
-\frac{k_\mu p_\nu-q_{1\mu}p_\nu-q_{1\mu}k_\nu}{((k-q_1)^2-m^2)
((p+k)^2-m^2)}]\nonumber \\
&=&-2im\Gamma_5^{\alpha\beta}-
\frac{1}{2\pi^2}\epsilon^{\mu\nu\alpha\beta}q_{1\mu}
q_{2\nu}[M_a(q_1;p,q_2)+M_a(q_2;p,q_1)].
\end{eqnarray}
After adding the measure contribution to this, we find:
\begin{eqnarray}\label{nid}
p_\mu\Gamma_5^{\mu\alpha\beta}&=&2m\Gamma_5^{\alpha\beta}-
\frac{i}{\pi^2}\epsilon^{\mu\nu\alpha\beta}q_{1\mu}
q_{2\nu}(M_a(p;q_1,q_2)+M_a(q_1;p,q_2)+M_a(q_2;p,q_1)).
\end{eqnarray}
\narrowtext
The extra piece exhibits current nonconservation and gives the
local limit (\ref{llim}), after the limit $M_a\rightarrow 1/6$ is taken.
We have seen that though we can maintain vector invariance in the
presence of axial couplings, anomalies appear in the axial sector.
We also note the presence of the fermion mass in the anomaly term.
This means that fermion doubling will only remove the anomaly in
the massless limit with the regulator on, but all mass terms are
suppressed by inverse powers of the nonlocal scale, so that in the
local limit even this dependence will disappear.

\subsection{Nonlocal Anomaly}

The analogous identities to those in Sect.\ref{local} may now be calculated,
and consistency with the perturbative expansion verified.
At this point, one of the main advantages of working with a theory
that is regulated explicitly at the Lagrangian level becomes
apparent, namely, the relevant transformations are already
regulated, and further considerations on defining and regulating
the measure are bypassed.
We can consider how the naive measure transforms under the
infinitesimal transformation in (\ref{nlocx}) to derive the
Jacobian order by order in the coupling constants.

Let us start by considering the local axial transformation:
\begin{eqnarray}\label{nlocx}
\delta\psi=-iE^2\omega\gamma^5[1-\bar{S}\Gamma]^{-1}\psi.
\end{eqnarray}
Under this transformation, the generating functional (\ref{genfun})
transforms to:
\widetext
\begin{equation}
Z[J_\mu,\bar{\eta},\eta]=\int d\mu[A^\mu,\psi,\bar{\psi}]exp(i\int
d^4x(L+\bar\psi\eta+\bar\eta\psi
+\delta L+\delta\bar\psi\eta+\bar\eta\delta\psi)+i\delta S_{meas}),
\end{equation}
where $\delta L$ is given by (\ref{deltaL}) and to third order in
couplings,
\begin{eqnarray}\label{deltas1}
\delta S_{meas}&=&\frac{i\Lambda^2}{2\pi^2}\int
\frac{dpdq}{(2\pi)^4}(2\pi)^2\delta^4(p+q)\omega(p)
[q^\mu B_\mu(q)M_v(q)+imC(q)M_c(q)]\nonumber \\
&+&\frac{1}{2\pi^2}\int\frac{dpdq_1dq_2}{(2\pi)^6}(2\pi)^4\delta^4(p+q_1+q_2)
M_a(q_1;p,q_2)\omega(p)\nonumber \\
&&\epsilon^{\mu\nu\alpha\beta}q_{1\mu}q_{2\nu}
[A_\alpha(q_1)A_\beta(q_2)+B_\alpha(q_1)B_\beta(q_2)],
\end{eqnarray}
where
\begin{equation}
M_c(q)=\int_0^\frac{1}{2}dt exp[t\frac{q^2}{\Lambda^2}
-\frac{1}{1-t}\frac{m^2}{\Lambda^2}].
\end{equation}
The full invariant measure we have constructed is invariant under
the transformation given in (\ref{delpsi}) and not (\ref{nlocx}).
This results in the $\delta S_{meas}$ term in (\ref{deltas1}), and ends up
giving the `anomalous' identity, since using (\ref{cond})
and (\ref{deltaL}), we get the weak operator identity:
\begin{eqnarray}\label{Js}
-ip_\mu J^\mu_5(p)&=&2imJ_5(p)+\delta S_{meas}=2imJ_5
+\frac{i\Lambda^2}{2\pi^2}M_v(p)p^\mu B_\mu(p)
-\frac{m\Lambda^2}{2\pi^2}C(p)M_c(p)\nonumber \\
&+&\frac{1}{2\pi^2}\int
\frac{dq_1dq_2}{(2\pi)^6}(2\pi)^4\delta^4(p+q_1+q_2)M_a(p;q_1,q_2)\nonumber \\
&&\epsilon^{\mu\nu\alpha\beta}q_{1\mu}q_{2\nu}
[A_\alpha(q_1)A_\beta(q_2)+B_\alpha(q_1)B_\beta(q_2)],
\end{eqnarray}
\narrowtext
resulting in the (respected) identity on the triangle graphs
(\ref{npcalc}).
Note that these identities apply only to Loop corrections generated by the
Lagrangian, we do not generate identities on contributions from the measure.
Operationally this is due to the fact that the presence of the measure does
not affect  the transformation (\ref{nlocax}), since we do not transform the
vector fields.
(The same result is true if we consider the vector analogue of (\ref{nlocax}).
Only if we consider the transformation properties of the photon as well, do we
derive the full Ward identities.)

The additional second order contributions to the current identity (\ref{Js})
remind us that the longitudinal degree of freedom of the axial vector field $B$
does not decouple from the $BB$ and $BC$ two-point functions (there is no
measure
contribution to these processes).
Strictly speaking, these are also anomalous terms, but in the local limit one
would introduce counterterms into the Lagrangian and absorb them into mass
redefinitions, effectively removing them from (\ref{Js}).
These terms do not contribute to the triangle anomaly.

Since we began with a regulated action and imposed vector invariance, the
anomaly
appeared (uniquely) in the axial sector.
We could also construct a conserved current from (\ref{Js}), however, as is
true in the local theories, we find that the Noether current does not
correspond to a conserved current in the quantum regime.

In contrast to other schemes, the anomaly relation here is
perturbative: it depends on a coupling constant series.
These higher order graphs are convergent and we expect
any additional contributions to vanish in the local limit,
reproducing the nonperturbative result of \cite{Fujikawa}.
We also expect that once renormalization is performed and the local
limit taken, there will be no correction to the anomalous
identity (\ref{llim}) from higher order loop corrections other than
those which contribute to the running of the coupling constant
\cite{adlerbardeen}.

\section{Chiral Gauge Invariance}

\subsection{Classical action}
We will now attempt to gauge the full chiral invariance in order to
explicitly demonstrate how the nonexistence of an invariant measure
in the nonlocal theory foils attempts to quantize it.
We begin with the chirally invariant local theory:
\begin{eqnarray}
L&=&{\bar\psi}(i\slashpart -m)\psi
-\frac{1}{4e^2}F^2_A-\frac{1}{4g^2}F^2_B
-{\bar\psi}\slashA\psi-{\bar\psi}\slashB\psi\nonumber \\
&\equiv &{\bar\psi} S^{-1}\psi+\frac{1}{2e^2}A^\mu
D^{-1}_{\mu\nu}A^\nu+\frac{1}{2g^2}B^\mu
D^{-1}_{\mu\nu}B^\nu\nonumber \\
&-&{\bar\psi}\slashA\psi-{\bar\psi}\slashB\gamma^5\psi,
\end{eqnarray}
possessing the gauge invariance:
\begin{eqnarray}\label{chinv}
\delta A^\mu&=&\theta^{,\mu},\quad\delta B^\mu=
\omega^{,\mu},\nonumber \\
\delta\psi&=&-i(\theta+\omega\gamma^5)\psi.
\end{eqnarray}
Introducing the shadow fields as before, we get
\begin{eqnarray}
L_{Sh}&=&{\bar\psi}\hat{S}^{-1}\psi+{\bar\phi}{\bar S}^{-
1}\phi+\frac{1}{2e^2}A^\mu D^{-1}_{\mu\nu}A^\nu+\frac{1}{2g^2}B^\mu
D^{-1}_{\mu\nu}B^\nu\nonumber \\
&-&({\bar\psi}+{\bar\phi})\slashA(\psi +\phi)
-({\bar\psi}+{\bar\phi})\slashB\gamma^5(\psi +\phi).
\end{eqnarray}
Integrating them out at the classical level (defining
$\Gamma=\slashA+\slashB\gamma^5$):
\begin{equation}
\phi=(1-\bar{S}\Gamma)^{-1}\bar{S}\Gamma\psi,
\end{equation}
\begin{eqnarray}
L_{NL}&=&\bar{\psi}\hat{S}^{-1}\psi+\frac{1}{2e^2}A^\mu\hat{D}^{-
1}_{\mu\nu}A^\nu+\frac{1}{2g^2}B^\mu\hat{D}^{-
1}_{\mu\nu}B^\nu\nonumber \\
&-&\bar{\psi}\Gamma(1-\bar{S}\Gamma)^{-1}\psi,
\end{eqnarray}
\begin{eqnarray}
\delta\psi&=&-iE^2(\theta+\omega\gamma^5)(\psi+\phi)\nonumber \\
&=&-iE^2(\theta+\omega\gamma^5)(1-\bar{S}\Gamma)^{-1}\psi,
\end{eqnarray}
with the current definitions in (\ref{nlocurrents}) still holding.

So far, this is all classical and there is no problem with it.
On attempting to quantize the theory using the path integral
formalism, we will discover that the invariant measure cannot be
generated by the method discussed in Sect. \ref{measure}, and therefore we
once more do not have a generating functional that generates graphs
respecting the classical symmetries.

\subsection{The Measure}

Building the measure as before, we easily derive the second order
pieces necessary to retain transversality of the vacuum
polarization:
\begin{eqnarray} \label{M2}
S_{meas}&=&-\frac{\Lambda^2}{4\pi^2}\int
\frac{dpdq}{(2\pi)^4}(2\pi)^4\delta^4(p+q)M_v^0(q)\nonumber \\
&\times&[A^\mu(p)A_\mu(q)+B^\mu(p)B_\mu(q)],
\end{eqnarray}
but now we run into problems at third order.
It is somewhat straightforward to show that the condition on the
measure at third order is given by:
\widetext
\begin{eqnarray}\label{dschir}
\delta S_{meas}&=&\frac{i}{2\pi^2}\int\frac{dpdq_1dq_2}{(2\pi)^6}
(2\pi)^4\delta^4(p+q_1+q_2)
\epsilon^{\mu\nu\alpha\beta}q_{1\mu}
M_a^0(q_1;p,q_2)\nonumber \\
&\times&[\delta B_\alpha(p)B_\nu(q_2)B_\beta(q_1)
+\delta B_\alpha(p)A_\beta(q_1)A_\nu(q_2)\nonumber \\
&+&\delta A_\alpha(p)A_\beta(q_1)B_\nu(q_2)
+\delta A_\alpha(p)B_\beta(q_1)A_\nu(q_2)].
\end{eqnarray}
\narrowtext
However, it is impossible to satisfy this condition without
introducing pole structure into the measure, or extra degrees of
freedom at the classical level.
First, one notices that the last two terms are identical to those
found in sect. \ref{measure}, and give rise to (\ref{Gams}).
The second term is also of the same order, and must be a
variation of the same action term.
It is not hard to check that the axial variation of (\ref{Gams})
gives two terms: one of the right form but the wrong sign, and the
other of the wrong form.
The $B^3$ piece produces a similar problem, in that there is no way
to construct a measure that gives just this one term.
In the local limit, nothing could survive anyway due to the
antisymmetry of the epsilon symbol.

As it stands, we cannot consistently quantize the theory and
maintain the axial symmetry, but there are a number of ways of
proceeding from here.
It is easily seen that we can write the measure as:
\begin{eqnarray}\label{pole}
S_{meas}&=&-\frac{1}{2\pi^2}\int\frac{dpdq_1dq_2}{(2\pi)^6}
(2\pi)^4\delta^4(p+q_1+q_2)
\epsilon^{\mu\nu\alpha\beta}\nonumber \\
&\times&M_a^0(q_1;p,q_2)\frac{p^\sigma}{p^2}B_\sigma(p)q_{1\mu}q_{2\alpha}
\nonumber \\
&\times&[B_\nu(q_2)B_\beta(q_1)+A_\nu(q_2)A_\beta(q_1)].
\end{eqnarray}
This expression explicitly introduces additional particle content into
the quantum regime, which can be identified as the longitudinal
degree of freedom of the axial boson due to the presence of the
longitudinal projector in (\ref{pole}), and so we also expect a
mass to be generated perturbatively \cite{grossj}.

Instead we could follow \cite{harada} and generate the Wess-Zumino
action through the introduction of a U(1) gauge parameter $\pi$
that accounts for the fact that the fermionic measure is not gauge
invariant, and therefore all gauge configurations must be summed
over.
This results in the additional integration over $\pi$ in the path
integral and the Wess-Zumino term:
\begin{eqnarray}
S_{W-Z}&=&-\frac{i}{2\pi^2}\int\frac{dpdq_1dq_2}{(2\pi)^6}
(2\pi)^4\delta^4(p+q_1+q_2)
\epsilon^{\mu\nu\alpha\beta}\nonumber \\
&\times&M_a^0(q_1;p,q_2)\pi(p)q_{1\mu}q_{2\alpha}\nonumber \\
&\times&[B_\nu(q_2)B_\beta(q_1)+A_\nu(q_2)A_\beta(q_1)].
\end{eqnarray}
This produces a gauge invariant generating functional, since under
$\pi\rightarrow\pi+\theta$ the action is invariant.

This, in fact, is merely a rewriting of the previous result.
Formerly we excited the longitudinal component of $B^\mu$ and thus
integrated over three degrees of freedom.
Here we excite only transverse $B^\mu$, but have an additional
field $\pi$, which can be identified with the longitudinal
component through a Stuekelberg-type construction.

Since we expect one to be produced perturbatively, we can also introduce an
axial boson mass into the classical Lagrangian:
\begin{equation}\label{mass}
L_M=\frac{1}{2}M^2B^\mu B_\mu
\end{equation}
explicitly breaking gauge invariance.
The Stuekelberg transformation then corresponds to performing a (finite)
inverse gauge transformation on all matter fields with the new auxiliary
field (in this case $\pi /M$) as the gauge field:
\begin{equation}
B^\mu\rightarrow B^\mu-\frac{\pi^{,\mu}}{M},\quad\psi\rightarrow
exp(i\frac{\pi}{M}\gamma^5)\psi,
\end{equation}
where $\delta\pi=M\omega$ supplements (\ref{chinv}) to recover
gauge invariance.
Only the mass term is non-invariant, and after gauge fixing is
performed (removing mixing terms between $B^\mu$ and $\pi^{,\mu}$)
we obtain the usual massive boson propagator in the Feynman gauge and the
auxiliary field Lagrangian:
\begin{equation}
L_\pi=\frac{1}{2}(\pi^{,\mu}\pi_{,\mu}-M^2\pi^2).
\end{equation}
At the classical level, we now have a massive axial boson, but
since the axial current is conserved, the longitudinal component
remains decoupled.
Nonlocalizing and quantizing, we then find that the condition
(\ref{dschir}) can now be satisfied with (\ref{Gams}) and,
\begin{eqnarray}
S_{meas}&=&-\frac{i}{2\pi^2}\int\frac{dpdq_1dq_2}{(2\pi)^6}
(2\pi)^4\delta^4(p+q_1+q_2)
\epsilon^{\mu\nu\alpha\beta}\nonumber \\
&\times&M_a^0(q_1;p,q_2)\frac{\pi(p)}{M}q_{1\mu}q_{2\alpha}\nonumber \\
&\times&[B_\nu(q_2)B_\beta(q_1)+A_\nu(q_2)A_\beta(q_1)].
\end{eqnarray}
This results in explicit coupling to the auxiliary field and hence the
longitudinal component of $B^\mu$.
It is a consistent measure and the Ward identities it generates
are satisfied, but the local limit is nonrenormalizable
\cite{grossj} and does not exist, and the massless limit also
does not exist.
We could apply the same method to the case where the fermion masses
are nonzero and essentially reproduce the results obtained in the earlier
sections, except that the measure would explicitly cancel off quadratic
divergences such as (\ref{vac}) in the axial sector as well.
The anomaly shows up here identically as in the 2-d Schwinger
model\cite{Hand}.

\section*{Conclusions}

Despite a claim to the contrary \cite{Moff-Wood}, the nonlocal regularization
of QED does `suffer' from an anomaly, and in the local limit
is consistent with the results in other schemes.
The advantage of the nonlocal formalism lies in the fact that the
currents of interest can be constructed in regulated form.
Any possible contributions from the measure needed to construct the
proper Ward identities may be derived directly, without resorting to
additional regularization or reequiring a `proper definition' of the measure.

\section*{Acknowledgements}

This work was supported by the Natural Sciences and Engineering
Research Council of Canada.
We thank Brian Hand for helpful discussions and for checking calculations.

\begin{figure}
\begin{center}
%
%
\bigphotons
\begin{picture}(20000,20000)
\drawline\photon[\E\REG](1000,10000)[6]
\put(-2000,\pfronty){$B_{\mu}$}
\drawarrow[\SE\ATTIP](\pmidx,\pmidy)
\global\advance\pmidy by 750
\global\advance\pmidx by -750
\put(\pmidx,\pmidy){\small$p$}
\drawline\fermion[\NE\REG](\photonbackx,\photonbacky)[4000]
\drawarrow[\SW\ATBASE](\pmidx,\pmidy)
\global\advance\pmidx by -950
\put(\pmidx,\pmidy){\small$k$}
\drawline\fermion[\SE\REG](\photonbackx,\photonbacky)[4000]
\drawarrow[\SE\ATBASE](\pmidx,\pmidy)
\global\advance\pmidx by -2500
\global\advance\pmidy by -1000
\put(\pmidx,\pmidy){\small$p+k$}
\drawline\fermion[\N\REG](\fermionbackx,\fermionbacky)[5656]
\drawarrow[\N\ATBASE](\pmidx,\pmidy)
\global\advance\pmidx by 550
\put(\pmidx,\pmidy){\small$k-q_{1}$}
\drawline\photon[\E\REG](\fermionbackx,\fermionbacky)[6]
\global\advance\pbackx by 950
\put(\pbackx,\pbacky){$A_{\alpha}$}
\drawarrow[\NW\ATBASE](\pmidx,\pmidy)
\global\advance\pmidy by 750
\global\advance\pmidx by 500
\put(\pmidx,\pmidy){\small$q_1$}
\drawline\photon[\E\REG](\fermionfrontx,\fermionfronty)[6]
\global\advance\pbackx by 950
\put(\pbackx,\pbacky){$A_{\beta}$}
\drawarrow[\NW\ATBASE](\pmidx,\pmidy)
\global\advance\pmidy by 750
\global\advance\pmidx by 500
\put(\pmidx,\pmidy){\small$q_2$}
\global\advance\pmidy by -5000
\global\advance\pmidx by -5000
\put(\pmidx,\pmidy){${-iA^{\mu\alpha\beta}}$}
\end{picture}
\\
\bigphotons
\begin{picture}(20000,20000)
\drawline\photon[\E\REG](1000,10000)[6]
\put(-2000,\pfronty){$B_{\mu}$}
\drawarrow[\SE\ATBASE](\pmidx,\pmidy)
\global\advance\pmidy by 750
\global\advance\pmidx by -750
\put(\pmidx,\pmidy){\small$p$}
\drawline\fermion[\NE\REG](\photonbackx,\photonbacky)[4000]
\drawarrow[\SW\ATBASE](\pmidx,\pmidy)
\global\advance\pmidx by -3500
\global\advance\pmidy by 750
\put(\pmidx,\pmidy){\small$-k-p$}
\drawline\fermion[\SE\REG](\photonbackx,\photonbacky)[4000]
\drawarrow[\SE\ATBASE](\pmidx,\pmidy)
\global\advance\pmidx by -1500
\global\advance\pmidy by -1000
\put(\pmidx,\pmidy){\small$-k$}
\drawline\fermion[\N\REG](\fermionbackx,\fermionbacky)[5656]
\drawarrow[\N\ATBASE](\pmidx,\pmidy)
\global\advance\pmidx by 550
\put(\pmidx,\pmidy){\small$q_{1}-k$}
\drawline\photon[\E\REG](\fermionbackx,\fermionbacky)[6]
\global\advance\pbackx by 950
\put(\pbackx,\pbacky){$A_{\beta}$}
\drawarrow[\NW\ATBASE](\pmidx,\pmidy)
\global\advance\pmidy by 750
\global\advance\pmidx by 500
\put(\pmidx,\pmidy){\small$q_1$}
\drawline\photon[\E\REG](\fermionfrontx,\fermionfronty)[6]
\global\advance\pbackx by 950
\put(\pbackx,\pbacky){$A_{\alpha}$}
\drawarrow[\NW\ATBASE](\pmidx,\pmidy)
\global\advance\pmidy by 750
\global\advance\pmidx by 500
\put(\pmidx,\pmidy){\small$q_2$}
\global\advance\pmidy by -5000
\global\advance\pmidx by -5000
\put(\pmidx,\pmidy){${-iB^{\mu\alpha\beta}}$}
\end{picture}

\end{center}
\caption{Graphs relevant for the calculation of the triangle Anomaly}
\label{triangle}
\end{figure}

\end{document}